\documentclass{kapproc} 

\RequirePackage{graphicx}%
\RequirePackage{epsf}%
\input{psfig.sty}

\upperandlowercase
\let\footnote\savefootnote
\let\footnotetext\savefootnotetext

\kluwerbib 

\begin{document}

\articletitle{Cluster Density and the IMF}

\chaptitlerunninghead{Cluster Density and IMF}

\author{Bruce G. Elmegreen}
\affil{IBM T.J. Watson Research Center, PO Box 218 Yorktown Hts.,
NY 10598, USA} \email{bge@watson.ibm.com}

 \begin{abstract}
Observed variations in the IMF are reviewed with an emphasis on
environmental density.  The local field IMF is not a dependable
representation of star-forming regions because of uncertain
assumptions about the star formation history.  The remote field
IMF studied in the LMC by several authors is clearly steeper than
most cluster IMFs, which have slopes close to the Salpeter value.
Local field regions of star formation, like Taurus, may have
relatively steep IMFs too. Very dense and massive clusters, like
super star clusters, could have flatter IMFs, or inner-truncated
IMFs. We propose that these variations are the result of three
distinct processes during star formation that affect the mass
function in different ways depending on mass range. At solar to
intermediate stellar masses, gas processes involving thermal
pressure and supersonic turbulence determine the basic scale for
stellar mass, starting with the observed pre-stellar
condensations, and they define the mass function from several
tenths to several solar masses. Brown dwarfs require
extraordinarily high pressures for fragmentation from the gas, and
presumably form inside the pre-stellar condensations during mutual
collisions, secondary fragmentations, or in disks. High mass stars
form in excess of the numbers expected from pure turbulent
fragmentation as pre-stellar condensations coalesce and accrete
with an enhanced gravitational cross section.  Variations in the
interaction rate, interaction strength, and accretion rate among
the primary fragments formed by turbulence lead to variations in
the relative proportions of brown dwarfs, solar to intermediate
mass stars, and high mass stars.  The observed IMF variations may
be explained in this way.
 \end{abstract}

to be published in ``IMF@50: A Fest-Colloquium in honor of
Edwin E. Salpeter,'' held at Abbazia di Spineto, Siena, Italy,
May 16-20, 2004. Kluwer Academic Publishers;
edited by E. Corbelli, F. Palla, and H. Zinnecker.

\section{Introduction}

The initial stellar mass function proposed by Edwin Salpeter in
1955 followed remarkably soon after the discovery of star
formation itself. Only several years earlier, Ambartsumian (1947)
noted that the clusters h and $\chi$ Persei and NGC 6231 were too
rarefied to resist galactic tidal forces. He said they should be
elongated by galactic shear, and because they were not, they had
to be young, expanding, and dispersing. Soon after, Blaauw (1952)
found the predicted 10 km s$^{-1}$ proper motions in the $\zeta$
Perseus association, and Zwicky (1953) proposed that stellar
expansion followed gas expulsion from bound newborn clusters. This
was the beginning of the recognition that stars had to form and
evolve continuously (Spitzer 1948; Hoyle 1953).

Salpeter (1955) reasoned that if stars form continuously, and if
their lifetimes depend on mass because of nuclear burning with a
steep mass-luminosity relation, then many more remnants from
massive stars than from low mass stars should populate the Milky
Way disk. Consequently, the stellar birth rate is a shallower
function of mass than the present day mass distribution.

The local field initial stellar mass function (IMF) derived by
Salpeter (1955) was based on the available population studies,
some of which were old even then, and on simple assumptions about
stellar lifetimes and galactic star formation history.  The result
was a power law initial mass function with a slope of
$\Gamma\sim-1.35$ (when plotted with equal intervals of $\log M$).
Considering the improvements in modern stellar data, there is no
reason to expect the mass function derived in 1955 would be the
same as today's. Indeed, subsequent studies almost always got
steeper field IMFs for the solar neighborhood: Scalo (1986) got
$\Gamma\sim-1.7$ at intermediate to high mass, and Rana (1987) got
$\Gamma\sim-1.8$ for $M>1.6$ M$_\odot$. The Salpeter value of
$\Gamma\sim-1.35$ predicts three times more massive stars
($10-100$ M$_\odot$) than intermediate mass stars ($1-10$
M$_\odot$) compared to the Scalo or Rana functions. Such an excess
can be ruled out for the local field today.

The steep slope of field IMFs becomes even more certain in recent
studies. Parker et al. (1998) derived $\Gamma=-1.80\pm0.09$ for
the LMC field that was far away from the HII regions catalogued by
Davies, Elliot \& Meaburn (1976).  Note the small statistical
error in this study. Massey et al. (1995, 2002) also surveyed the
remote field in the LMC and SMC: at distances greater than $30$ pc
from Lucke \& Hodge (1970) or Hodge (1986) OB associations, in a
survey complete to 25 M$_\odot$, $\Gamma\sim-3.6$ to $-4$ for a
constant star formation rate during the last 10 My.  There were
450 stars in the most recent Massey et al. LMC sample, which would
give a statistical uncertainty in the slope of only $\pm0.15$
(Elmegreen 1999).

These steep field IMF slopes are reminiscent of that found by
Garmany et al. (1982). In a survey of the solar neighborhood out
to 2.5 kpc, and complete for $M>20$ M$_\odot$, Garmany et al.
found $\Gamma=-1.6$ overall, $\Gamma\sim-1.3$ inside the Solar
circle, and $\Gamma\sim-2.1$ outside the Solar circle. They
proposed that the difference between these values arose because of
an excess of massive stars in the associations of the Carina and
Cygnus spiral arms, which are inside the Solar circle.  That is,
the low density regions have steep IMF slopes and the high density
regions have shallow IMF slopes. This observation led to the
concept of bimodal star formation: G\"usten \& Mezger (1983)
proposed that metallicity gradients in the Galaxy arose from an
interarm IMF that was steeper than the spiral arm IMF. Larson
(1986) proposed that a sub-population formed with a shallow IMF
would have more remnants contributing to dark matter. Such extreme
bimodality was never confirmed, although the steep field IMF slope
persisted. Also, in contrast to Garmany et al., a recent study of
the Milky Way disk by Casassus et al. (2000) got the same IMF
inside and outside the Solar circle using IR sources in
ultra-compact HII regions. They got the steep value in both
places, however, $\Gamma\sim-2$.

The problem with most of the field IMFs is that they are subject
to systematic uncertainties in the essential assumptions: the star
formation history, the mass dependence of the galactic scale
height, galactic radial migration, etc.. Fits to a constant
historical star formation rate in the Milky Way are inconsistent
with the Madau et al. (1996) result, for example, which suggests
that cosmological star formation was significantly higher 10 Gy
ago, the likely age of the Milky Way disk (this excess could be
from different galaxies, however). Also uncertain is the disk
formation history, considering the possibility of gaseous
accretion and minor mergers. Similar uncertainties arise for
recent times: a burst of star formation from the most recent
passage of a spiral density wave would give a decaying star
formation rate locally, changing the history corrections in the
IMF and making the slope shallower for $M>5$ M$_\odot$. In fact it
is likely that local star formation varies on a 100 My time scale.
These uncertainties translate into unknown corrections for the IMF
during the conversion from present day star counts to relative
fractions at birth. For this reason, moderately steep field IMFs
at intermediate to high mass may not be representative of the
average IMF in field regions of star formation.

The extreme fields in the Large and Small Magellanic Clouds seem
different, however. The steep slope differs from the Salpeter
value by a statistically significant margin, there is no scale
height uncertainty because the line of sight integrates through
the entire disk, massive stars are easily detected, and their 10
My lifetimes minimize uncertainties in the star formation history.
The resulting IMF seems trustworthy.  In addition, small-scale
local star formation, as in Taurus, could be unusually steep too
for $M>1$ M$_\odot$ (Luhman 2000).

Cluster IMFs are difficult to measure because of the small number
of stars in most clusters (Scalo 1998) and because of mass
segregation (de Grijs et al. 2002).  Some clusters have the same
steep IMF as the field (e.g., the upper Sco OB association --
Preibisch et al. 2002), but other clusters have what appears to be
an excess of high mass stars in certain subgroups, making their
overall slopes shallower. W51 (Okumura et al. 2000) is an example
where the intermediate mass IMF has a slope of $\sim-1.8$, but
sub-regions 2 and 3 have a statistically significant excess of
stars at $M\sim60$ M$_\odot$ (a 2 to 3 $\sigma$ deviation).

After uncertain corrections for mass segregation, field
contamination, and completeness, most clusters have IMF slopes
that are significantly shallower than $\Gamma=-1.8$, and more like
the original Salpeter value of $\Gamma=-1.35$. R136 in the 30 Dor
region of the LMC is a good example (Massey \& Hunter 1998). Other
examples are h and $\chi$ Persei (Slesnick, Hillenbrand \& Massey
2002), NGC 604 in M33 (Gonz\'alez Delgado \& Perez 2000), NGC 1960
and NGC 2194 (Sanner et al. 2000), and NGC 6611 (Belikov et al.
2000). Massey \& Hunter (1998) proposed that the Salpeter IMF
spans a factor of 200 in cluster density.  If this is the case,
then clustered star formation produces a shallower IMF than
extreme field star formation.

The most extreme cases of clustered star formation are in
starburst regions (Rieke et al. 1993), particularly in super star
clusters. Sternberg (1998) derived a high $L/M$ for the super star
cluster NGC 1705-1, and inferred that either $|\Gamma|<1$ or there
is an inner-mass cutoff greater than the local value of $0.5$
M$_\odot$ (Kroupa 2001).  Smith \& Gallagher (2001) got a high
$L/M$ in M82F and proposed an inner cutoff of 2 to 3 M$_\odot$ for
$\Gamma=-1.3$. They also confirmed the Sternberg result for NGC
1705-1.  Alonso-Herrero et al. (2001) got a high $L/M$ in the
starburst galaxy NGC 1614.  F\"orster Schreiber et al. (2003)
found the same for M82, proposing an inner IMF cutoff of 2 to 3
M$_\odot$ for $\Gamma=-1.35$. Similarly, McCrady et al. (2003)
found a deficit in low mass stars in the cluster MGG-11 of M82.
Not all super star clusters require an inner IMF cutoff: N1569-A
(Sternberg 1998), NGC 6946 (Larsen et al. 2001), and M82: MGG-9
(McCrady et al. 2003) do not.

We might summarize these results as follows: The field IMF is
systematically steeper than the cluster IMFs, but both are
uncertain. Nevertheless, in the field and in low-density clustered
regions, the IMF slope is fairly steep, perhaps $\Gamma=-1.8$ or
steeper, while in clusters it is more shallow, $\Gamma=-1.3$ or
shallower, with even more of a high mass bias for the most extreme
clusters in starburst regions.  This IMF difference suggests a
difference in star formation mechanisms, and is reminiscent of
Motte \& Andr\'e's (2001) suggestion that accretion processes and
pre-stellar condensation sizes are different in the low density
regions of Taurus compared to the higher density associations in
Perseus and Ophiuchus. Significantly steep IMFs in some dispersed
associations, and very steep IMFs in remote fields contribute to
this picture.  In the disks of low surface brightness galaxies, a
high mass-to-light ratio suggests the entire IMF is steep with
$\Gamma=-2.85$ (Lee et al. 2004); star formation in the
low-density ``field'' mode could be pervasive.

In addition to these observations, the segregation of stellar mass
in many clusters, apparently at birth (Bonnell \& Davies 1998),
also suggests high mass stars prefer dense environments (Pandey,
Mahra \& Sagar 1992; Subramaniam, Sagar \& Bhatt 1993; Malumuth \&
Heap 1994; Brandl et al. 1996; Fischer et al. 1998; Hillenbrand \&
Hartmann 1998; Figer, McLean, \& Morris 1999; Le Duigou \&
Kn\"odlseder 2002; Stolte et al. 2002; Sirianni et al. 2002;
Muench et al. 2003; Gouliermis et al. 2004; Lyo et al. 2004).

We conclude that dense regions favor massive star formation. A
more comprehensive survey of the observations is in Elmegreen
(2004).

\section{Theoretical expectations}

There has long been a notion that massive star formation should be
more likely in dense environments. It takes ultra-high pressures
to confine the winds and radiation from massive stars (Garay \&
Lizano 1999; Yorke \& Sonnhalter 2002; Churchwell 2002; McKee \&
Tan 2003), and these pressures require high density cloud cores.
Massive stars could also form by enhanced accretion from high
density gas reservoirs (Zinnecker 1982; Larson 1999, 2002; Myers
2000; Bonnell et al. 1997, 2001, 2004), by the coalescence of
pre-stellar condensations (Zinnecker 1986; Larson 1990; Price \&
Podsiadlowski 1995; Stahler, Palla \& Ho 2000), coalescence after
accretion drag (Bonnell, Bate, \& Zinnecker 1998), or coalescence
after accretion-induced cloud-core contraction (Bonnell, Bate \&
Zinnecker 1998; Bonnell \& Bate 2002).  Indeed, numerical
simulations in Gammie et al. (2003) showed that the high-mass part
of the IMF gets shallower with time as a result of coalescence and
enhanced accretion. Li, et al. (2004) also found enhanced
accretion and coalescence played a role in massive star formation.

Nearly every attempt to confirm these ideas has been incomplete or
fraught with selection effects. The observation of larger
most-massive stars in denser clusters would support this picture
(Testi, Palla, \& Natta 1999), but it appears instead to be the
result of sampling statistics: all of the clusters in their survey
were about the same size and so the cluster density correlated
with total cluster mass (Bonnell \& Clarke 1999; Elmegreen 2000).
More massive clusters usually give more massive most-massive stars
by sampling further out in the IMF. Earlier observations of
cloud-mass versus star-mass correlations (Larson 1982) apparently
had the same size of sample effect (Elmegreen 1983).

\begin{figure}[tb]
\centerline{\psfig{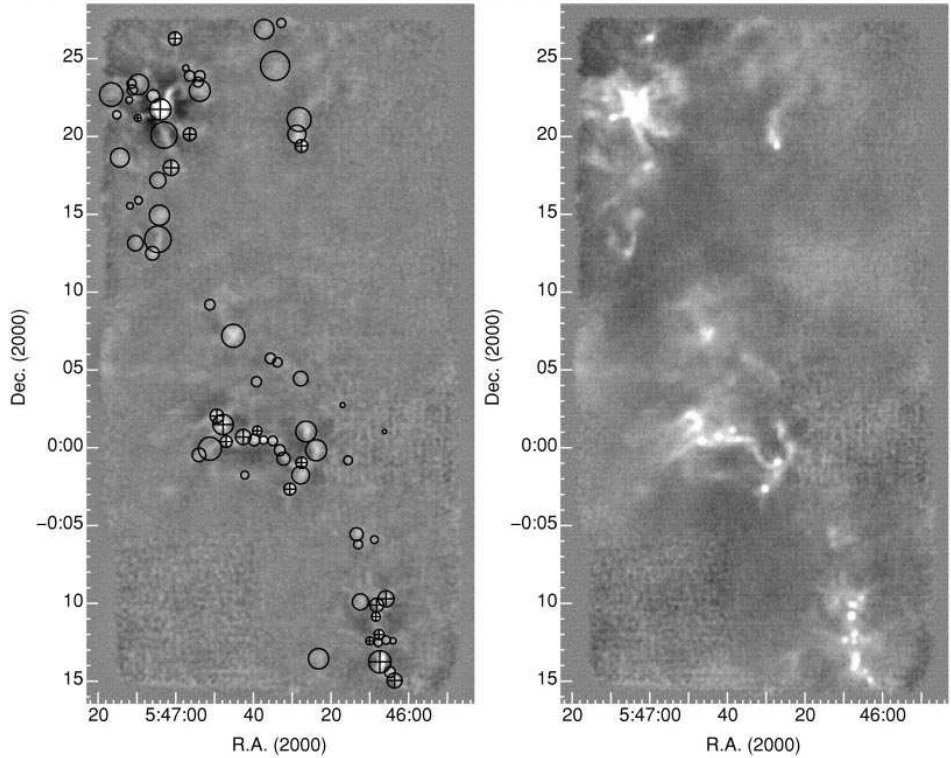}} \caption{850 $\mu$m
continuum observations of Orion from Johnstone et al. (2001). The
left panel shows identifications of pre-stellar condensations made
by a clump-finding algorithm, and the right panel shows the
emission. These condensations are typical for dense pre-stellar
objects. In this cluster-forming environment, they are often close
together, suggesting that some will coalescence in the future.
Reproduced by permission of the AAS.}
\end{figure}

We are concerned with a more subtle effect here however, and one
dependent on both density and mass rather than just mass. The
observations suggest all clusters are more or less the same, i.e.,
having a fairly shallow IMF, so differences between clusters will
be dominated by sampling, selection, and mass segregation.
However, star formation at very low density, in the extreme field
or in low surface brightness galaxies, is apparently different,
producing steep IMFs. We choose to emphasize here the difference
between clusters as a whole and the remote field, rather than a
density dependence for the IMF among the clustered population
alone. With this more extreme comparison, it seems reasonable to
think that the relatively isolated star formation of remote field
regions will not produce much coalescence of pre-stellar
condensations, and will therefore lack the ``cluster-mode'' of
competitive accretion and dense core interactions that are
illustrated by numerical simulations (e.g., Bonnell, Bate, \& Vine
2003). In this limited sense, we believe the massive-star IMF
should vary with density (Elmegreen 2004; Shadmehri 2004).

Figure 1 shows the mm-wave continuum emission in the Orion B
region, from Johnstone et al. (2001). The tiny knots of emission
are examples of pre-stellar condensations, similar to those found
in many other regions (Motte et al. 1998). The figure shows how
closely packed many of these condensations are, giving credence to
the idea that some will interact or coalesce with their nearest
neighbors.  A statistical study of such coalescence, limited to
the pre-stellar condensation phase when the objects are fairly
large ($\sim10^4$ AU), suggests that the most massive and dense
clusters should have the most coalescence (Elmegreen \& Shadmehri
2003).   That is, the ratio of the coalescence time to the
collapse time of pre-stellar condensations decreases for more
massive clusters or for denser clusters.  The mass dependence
arises because more massive clusters have more massive stars,
which are more strongly attracting to other pre-stellar
condensations.   The pre-stellar condensations are more widely
separated in Taurus (Motte \& Andr\'e 2001) than in Orion B. This
supports our view that interactions at this phase are relatively
less important in the low-density field environment.

Theoretical considerations in Elmegreen (2004) and  Shadmehri
(2004) suggest there are three distinct regimes of physical
processes in the IMF:
\begin{itemize}
\item For solar to intermediate mass stars, cloud or gas processes
connected with turbulence and the thermal Jeans mass, $M_{J0}$,
are important for the formation of pre-stellar condensations
(e.g., see review in Mac Low \& Klessen 2004, and see Gammie et
al. 2003; Li, et al. 2004).

\item Brown dwarfs differ because gravitational instabilities at
such low mass require ultra-high pressures.  These naturally occur
inside the $M_{J0}$ pieces formed by cloud processes; i.e., in
disk instabilities surrounding $M_{J0}$ protostars, in the ejecta
from collisions between $M_{J0}$ objects, in the early ejection of
accreting protostars from tight clusters, and so on, as in the
usual models (Padoan \& Nordlund 2002; Reipurth \& Clarke 2001;
Bate, Bonnell \& Bromm 2002; Preibisch et al. 2003; Kroupa \&
Bouvier 2003). In addition, collisions between $M_{J0}$ objects
should induced gravitational instabilities in the shocked gas.
Because the pre-stellar condensation density is higher than the
ambient cloud density by a factor of $\sim100$, the Jeans mass for
these shock instabilities will be lower than $M_{J0}$ by a factor
of $\sim10$, placing them in the brown dwarf range.

\item High mass stars form by cloud processes too, but their
formation rate can be greatly enhanced by the coalescence of
$M_{J0}$ pieces and by gas accretion.  These are runaway processes
considering gravitationally enhanced cross sections, and so become
more prominent when the condensation mass exceeds $M_{J0}$ by a
factor of $\sim10$.

\end{itemize}

\begin{figure}[tb]
\centerline{\psfig{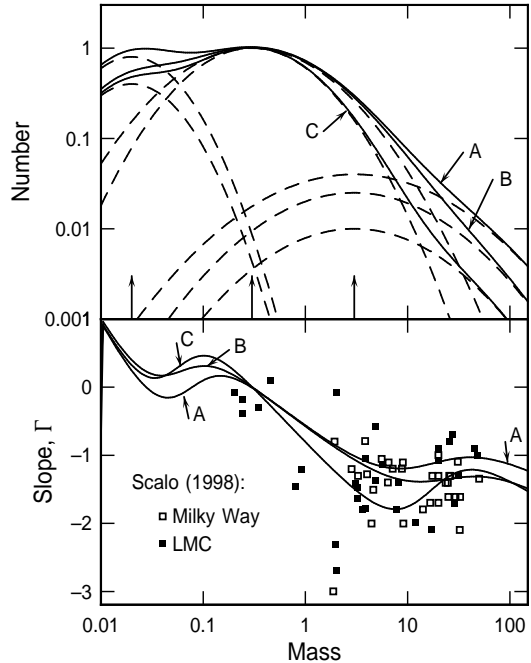}} \vspace{-10mm}
\caption{Three component models of the IMF with the distinct
components indicated by dashed lines. The top panel shows the IMFs
and the bottom panel shows the slopes along with observations from
Scalo (1998). Each component is a log-normal with a characteristic
amplitude, central mass (indicated by arrows along the abscissa of
the top panel), and dispersion. Curves A, B, and C correspond in
the top and bottom panels. The combined IMFs have pseudo-power
laws at intermediate to high mass. Based on a figure in Elmegreen
(2004).}
\end{figure}

We consider that these three regimes of star formation produce
three separate IMFs that usually combine into one in a way that
gives the seemingly universal power law with a low mass turnover
at about $M_{J0}$.   However, variations in the importance of
these three processes, particularly with variations in the ambient
cloud density, produce variations in the relative amplitudes of
the three IMFs, and these variations have the effect of changing
the slope of the power-law fit at intermediate to high mass.

Such variations can also change the proportion of brown dwarfs and
normal stars. IC 348 (Preibisch, Stanke \& Zinnecker 2003; Muench
et al. 2003; Luhman et al. 2003) and Taurus (Luhman 2000;
Brice\~no et al. 2002) have brown dwarf-to-star ratios that are
$\sim2$ times lower than in many other local clusters, including
the Orion trapezium cluster (Hillenbrand \& Carpenter 2000; Luhman
et al. 2000; Muench et al. 2002), the Pleiades (Bouvier et al.
1998; Luhman et al. 2000), M35 (Barrado y Navascu\'es et al. 2001)
and the galactic field (Reid et al. 1999). IC 348 and Taurus
differ even in the subsolar range (Luhman et al. 2003).

Figure 2 shows an example of how variability in three distinct
mass intervals, separated by factors of 10, can produce
variability in the summed IMF that is similar to what is observed
(from Elmegreen 2004). Three distinct log-normals, one for each
physical process, are shown to sum to a near power law between 1
and 100 M$_\odot$. The intermediate-to-high mass slope gets
shallower as the high-mass contribution increases. The brown dwarf
range can be made to vary too.

Figure 3 shows IMF models where a cloud forms stars with a locally
log-normal mass distribution that has a central mass and
dispersion increasing with cloud density (from Elmegreen 2004).
The cloud density has the form
$\rho_c(r)=\left(1+\left[r/r_0\right]^2\right)^{-1}$ for core
radius is $r_0$. The local IMF is taken to be
$f(M)=A\exp\left(-B\left[\log\left\{M/M_0\right\}\right]^2\right)$
for exponential factor $B=B_1-B_2\rho_c(r)$ and central mass
$M_0=M_1+M_2\rho_c(r)$. With these expressions, the local
log-normal is broader and shifted toward higher mass in the cloud
core.  The Miller-Scalo (1979) IMF has $B_1=1.08$ and $M_1=0.1$
M$_\odot$ with no density dependence. The total mass function in
the cloud is determined by integrating over radius out to
$r_{max}$ with a weighting factor equal to the $3/2$ power of
density; this accounts for the available mass and a star formation
rate locally proportional to the dynamical rate.  The figure shows
the Miller-Scalo IMF ($B_2=M_2=0$) as a dotted line and sample
IMFs with $B_2=B_1$, $M_2=1,$ and $r_{max}=2r_0$ (solid line),
$5r_0$ (dashed), and $10r_0$ (dot-dashed). The IMF slope is
shallower for smaller $r_{max}$, indicating mass segregation. The
bottom panel plots mean separation between the logs of the masses
in the IMF along with observations of R136 from Massey \& Hunter
(1998). The slope of the distribution of points is the negative
power, $-\Gamma$, of a power-law IMF.  The R136 points have
$\Gamma\sim-1.1$.  Clearly a variable IMF can be made to fit this
observation using only local IMFs that are log-normal in form.

\begin{figure}[tb]
\centerline{\psfig{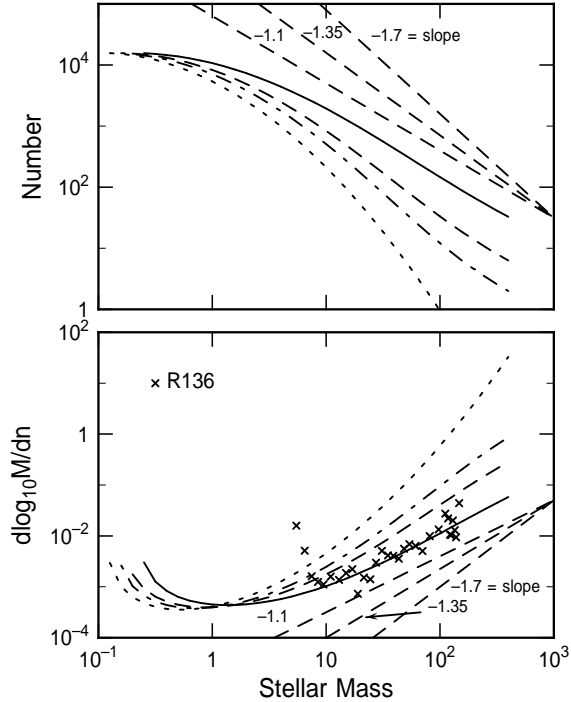}}\vspace{-5mm}
\caption{(top) IMF model based on a log-normal mass distribution
in which the dispersion increases with density. The IMF is
integrated over a cloud density profile out to 2, 5, and 10 cloud
core radii for solid, dashed, and dot-dashed lines. The dotted
line is the Miller-Scalo (1979) IMF. (bottom) The mean separation
between the log of the masses for the model IMFs shown in the top
and for the R136 cluster in the LMC.}
\end{figure}

\section{Summary}

There is apparently no ``Universal IMF.'' Low-pressure star
formation in remote field regions is characterized by
widely-separated, pre-stellar condensations with masses centered
on the thermal Jeans mass, ($M_{J0}$). These condensations rarely
interact, so they produce a ``gas-only'' IMF with no high-mass
component and a steep slope at intermediate to high stellar mass.
High-pressure star formation in clusters produces the same
gas-only IMF for pre-stellar condensations, but these
condensations collapse and collide to make ultra-high pressure
regions (disks, shocks, etc.), leading to brown dwarfs. The same
high-pressure cluster environment also promotes coalescence of
pre-stellar condensations to build up more massive stars.  Thus at
least one process of brown dwarf formation correlates with the
enhanced formation of massive stars.

Stellar IMFs for galaxies and clusters are about the same because
most stars form in clusters.  There is apparently little
correlation (aside from sampling effects) between stellar mass and
cluster mass for these stars.  At very low densities and
pressures, as in the field, or in low surface brightness galaxies,
or in some dwarf galaxies, the IMF should be relatively steep
($\Gamma\le-1.7$), with few massive stars because of a lack of
significant pre-main-sequence coalescences, and relatively few
brown dwarfs because of a lack of collision remnants.  At very
high densities, as in dense clusters, the IMF should be shallow,
like the Salpeter IMF or perhaps shallower, because of the
coalescence of pre-stellar condensations and because of enhanced
gas accretion. Brown dwarfs should be relatively common in these
regions too because of the high pressures formed by interactions.

{}
\end{document}